\documentclass[11pt]{article}

\usepackage[top=1in,bottom=1in,left=1in,right=1in]{geometry}
\usepackage{natbib}
\usepackage[unicode=true,bookmarks=true,bookmarksnumbered=false,bookmarksopen=false,breaklinks=false,pdfborder={0 0 1},backref=false,colorlinks=true]{hyperref}
\hypersetup{citecolor=blue}
\usepackage{url}

\usepackage{amsmath}
\usepackage{mathrsfs}
\usepackage{amssymb}
\usepackage{amsthm}
\usepackage{mathtools,amssymb}
\usepackage[normalem]{ulem}

\usepackage{float}
\usepackage{rotfloat}

\floatstyle{ruled}
\newfloat{algorithm}{tpb}{loa}
\providecommand{\algorithmname}{Algorithm}
\floatname{algorithm}{\protect\algorithmname}

\usepackage{algpseudocode,algorithm}
\algnewcommand\algorithmicinput{\textbf{Input}:}
\algnewcommand\algorithmicoutput{\textbf{Output}:}
\algnewcommand\INPUT{\item[\algorithmicinput]}
\algnewcommand\OUTPUT{\item[\algorithmicoutput]}

\usepackage{graphicx}
\usepackage[caption=false,labelformat=simple]{subfig}

\graphicspath{{/Users/yizhao/Dropbox/MyFolder/Biostat-IU/Projects/Lexin/2019/HD-Cause/}}

\usepackage{array}
\newcolumntype{L}[1]{>{\raggedright\let\newline\\\arraybackslash\hspace{0pt}}m{#1}}
\newcolumntype{C}[1]{>{\centering\let\newline\\\arraybackslash\hspace{0pt}}m{#1}}
\newcolumntype{R}[1]{>{\raggedleft\let\newline\\\arraybackslash\hspace{0pt}}m{#1}}

\usepackage{rotating}
\usepackage{multirow}

\usepackage{tikz}
\usetikzlibrary{shapes,arrows,backgrounds,calc,positioning,fit,petri,plotmarks}
\usetikzlibrary{arrows}

\usepackage{enumerate}
\usepackage{paralist}

\newcommand*{\affaddr}[1]{#1} 
\newcommand*{\affmark}[1][*]{\textsuperscript{#1}}

\global\long\def\bX{\mathbf{X}}
\global\long\def\bx{\mathbf{x}}
\global\long\def\bY{\mathbf{Y}}

\global\long\def\ba{\mathbf{a}}

\global\long\def\bD{\mathbf{D}}

\global\long\def\bw{\mathbf{w}}

\global\long\def\bV{\mathbf{V}}

\global\long\def\bM{\mathbf{M}}

\global\long\def\balpha{\boldsymbol{\alpha}}
\global\long\def\bbeta{\boldsymbol{\beta}}
\global\long\def\bgamma{\boldsymbol{\gamma}}
\global\long\def\bepsilon{\boldsymbol{\epsilon}}
\global\long\def\bdeta{\boldsymbol{\eta}}

\global\long\def\bmu{\boldsymbol{\mu}}
\global\long\def\bnu{\boldsymbol{\nu}}
\global\long\def\btau{\boldsymbol{\tau}}

\title{Multimodal Data Integration via Mediation Analysis with High-Dimensional Exposures and Mediators}

\author{%
    Yi Zhao\affmark[1], Lexin Li\affmark[2], for the Alzheimer's Disease Neuroimaging
Initiative\footnote{Data used in preparation of this article were obtained from the Alzheimer's Disease Neuroimaging Initiative (ADNI) database (\url{adni.loni.usc.edu}). As such, the investigators within the ADNI contributed to the design and implementation of ADNI and/or provided data but did not participate in analysis or writing of this report. A complete list of ADNI investigators can be found at: \url{http://adni.loni.usc.edu/wp-content/uploads/how_to_apply/ADNI_Acknowledgement_List.pdf}} \\
    \affaddr{\affmark[1]Department of Biostatistics and Health Data Science, Indiana University School of Medicine} \\
    \affaddr{\affmark[2]Department of Biostatistics and Epidemiology, University of California, Berkeley}
}

\date{}

\providecommand{\keywords}[1]
{
  {\small 
  \textbf{Keywords:} #1}
}

\begin{document}

\maketitle

\thispagestyle{empty}

\begin{abstract}
Motivated by an imaging proteomics study for Alzheimer's disease (AD), in this article, we propose a mediation analysis approach with high-dimensional exposures and high-dimensional mediators to integrate data collected from multiple platforms. The proposed method combines principal component analysis with penalized least squares estimation for a set of linear structural equation models. The former reduces the dimensionality and produces uncorrelated linear combinations of the exposure variables, whereas the latter achieves simultaneous path selection and effect estimation while allowing the mediators to be correlated. Applying the method to the AD data identifies numerous interesting protein peptides, brain regions, and protein-structure-memory paths, which are in accordance with and also supplement existing findings of AD research. Additional simulations further demonstrate the effective empirical performance of the method. 
\end{abstract}




\clearpage
\setcounter{page}{1}

\section{Introduction}
\label{sec:intro}

Alzheimer's disease (AD) is an irreversible neurodegenerative disorder and is characterized by progressive impairment of cognitive and bodily functions and ultimate death. It is currently affecting over 5.8 million American adults aged 65 years or older. Meanwhile, its prevalence continues to grow and is projected to reach 13.8 million by 2050 \citep{AD2020}. Multimodal technologies have transformed AD research in recent years, by collecting different types of data from the same group of subjects and enabling the investigation of complex interrelated mechanisms underlying AD development. Notable examples include multimodal neuroimaging studies of the joint impact of brain structure and function on the disorders \citep{liu2015multimodal, higgins2018integrative}, and imaging genetics studies of the impact of genetic variants on the brain then the disease outcome \citep{nathoo2019review}, among others. 

Our motivation is an imaging proteomics study, which is part of the Alzheimer's Disease Neuroimaging Initiative (ADNI) that aims to identify biomarkers for early detection and tracking of AD and to assist the development of prevention and intervention strategies. Amyloid-$\beta$ is a microscopic brain protein fragment, denotes peptides of 36 to 43 amino acids, and is part of a larger protein called amyloid precursor protein. Tau is a group of microtubule-associated proteins predominantly found in brain cells and performs the function of stabilizing microtubules. Amyloid-$\beta$ is the main component of amyloid plaques, while tau is the main component of neurofibrillary tangles, both of which are commonly found in the brains of AD patients. Models of AD pathophysiology hypothesize a temporal sequence, in which accumulations of amyloid-$\beta$ plaques and neurofibrillary tangles disrupt cell-to-cell communications and destroy brain cells, leading to brain structural atrophy in regions such as the hippocampus, and ultimately a clinical decline in cognition  \citep{mormino2009episodic}. However, it remains unclear how these two proteins interact with each other and with other proteins in the cerebrospinal fluid (CSF), and how those proteins together subsequently affect brain atrophy and disease progression. In our study, we aim to investigate simultaneously the interrelations of multiple protein peptides in the CSF, along with multiple brain regions of the whole brain, and their impact on memory. 

The problem can be formulated as a mediation analysis, where the goal is to identify and explain the mechanism, or path, that underlies an observed relationship between an exposure and an outcome variable, through the inclusion of an intermediate variable known as a mediator. It decomposes the effect of exposure on the outcome into a direct effect and an indirect effect, the latter of which indicates whether the mediator is on a path from the exposure to the outcome. In our multimodal AD study, the measurements of the amount of multiple protein peptides serve as the exposure variables, the volumetric measurements of multiple brain regions serve as the potential mediators, and a composite memory score serves as the outcome. See Section~\ref{sec:motivation} for more details about the study and the data. Our objective is to identify paths from proteins to brain regional atrophies that lead to memory decline.

Mediation analysis was first proposed with a single exposure and a single mediator \citep{Baron1986}. See \citet{VanderWeele2016} for a review of mediation analysis and many references therein. In our setting, both the exposure variables and mediators are multivariate and potentially high-dimensional. While there have been numerous extensions of mediation analysis to account for multiple mediators \citep[see, e.g.,][among many others]{zhao2016pathway, chen2017high, song2018bayesian}, there have been very few works studying multivariate exposures, or both multivariate exposures and mediators. Recently, \citet{zhang2019high, aung2020application, long2020framework} proposed new approaches for mediation analysis of multivariate exposures and mediators. In particular, \citet{zhang2019high} developed two regularization procedures and applied them to a mouse f2 dataset for diabetes, taking SNP genotypes as the exposures, islet gene expressions as the mediators, and insulin level as the outcome. However, they required the mediators to be independent, which hardly holds in our setting, as different brain regions are generally believed to influence each other. \citet{aung2020application} studied environmental toxicants on pregnancy outcomes, taking toxicants as the exposures, endogenous biomarkers such as inflammation and oxidative stress as the mediators, and gestational age at delivery as the outcome. A key strategy of their analysis was to reduce the exposure dimension by creating environmental risk scores for a small number of groups based on the domain knowledge. They showed that the between-group correlation in the reduced exposures is negligible. However, such prior domain knowledge may not always be available. \citet{long2020framework} proposed a general mediation framework to identify proteins that mediate the effect of metabolic gene expressions on survival for a type of kidney cancer, taking mRNA levels as the exposures, protein measures as the mediators, and survival time as the outcome. Nevertheless, they implicitly required the dimensions of the exposures and mediators cannot be too high, and thus their method is not directly applicable to our setting, where the number of exposures and mediators can both be potentially larger than the sample size. 

In this article, we propose a mediation analysis approach, with both high-dimensional exposures and high-dimensional mediators, for multimodal data analysis. The method integrates principal components analysis (PCA) with penalized least squares estimation for a set of linear structural equation models. The former reduces the dimensionality and produces uncorrelated linear combinations of the exposure variables, whereas the latter achieves path selection and effect estimation while allowing the multivariate mediators to be potentially correlated. We apply this approach to the imaging proteomics study of AD to integrate CSF proteomics, brain volumes, and a memory measure of mild cognitive impairment (MCI) subjects in ADNI. We identify several interesting protein peptides, brain regions, and protein-structure-memory paths that are in accordance with and also supplement the existing knowledge of AD. Additional simulations further demonstrate the efficacy of the method. Similar to \citet{zhang2019high, aung2020application, long2020framework}, our approach is among the first attempts to conduct mediation analysis where both the exposures and mediators are high-dimensional. But unlike the existing solutions, we do not restrict the dimensionality or the correlation structures and do not require additional domain knowledge of the exposures or mediators. Moreover, although focusing on a multimodal neuroimaging study in this article, our proposed method is equally applicable to a wide range of multimodal data integration problems, e.g., the multi-omics data analysis \citep{Richardson2016}, and the multimodal healthcare study \citep{Cai2019}. As such, our proposal makes a useful addition to the general toolbox of both mediation analysis and multimodal data integration. 
 
The rest of the article is organized as follows. Section~\ref{sec:motivation} introduces the motivating imaging proteomics data of AD. Section~\ref{sec:model} presents the proposed model and estimation approach. Section~\ref{sec:ADNI} analyzes the AD dataset, with a detailed discussion on the identified protein peptides, brain regions, and path. Section~\ref{sec:sim} complements with additional simulation results to demonstrate the empirical performance of the method. 

\section{Motivating example}
\label{sec:motivation}

While Alzheimer's disease is becoming a major public health challenge as the population ages, there is no effective treatment for AD that is capable of stopping or slowing the associated cognitive and neuronal degradation. Therefore, understanding the disease pathology, identifying biological markers, and finding early diagnosis and intervention strategies are of critical importance \citep{AD2020}. Among numerous AD-related proteins in the CSF, amyloid-$\beta$ and tau are two major proteins that are consistently identified in the brains of AD patients, and their abnormal abundance generally indicates AD pathology \citep{Jagust2018}. Even though there has been evidence suggesting a pathological connection between amyloid-$\beta$ deposition, hippocampus atrophy, and memory decline \citep{mormino2009episodic}, it remains largely unknown how amyloid-$\beta$ and tau interact with each other, how they interact with other proteins in the CSF, and how these proteins together affect the downstream brain atrophy and cognitive outcome. In our study, we aim to delineate the regulatory relationships among multiple CSF proteins, structural atrophy of the whole brain, and cognitive behavior, and to identify important biological paths. 

The data used in our study are obtained from the Alzheimer's Disease Neuroimaging Initiative (ADNI, \url{adni.loni.usc.edu}). The CSF proteomics data were obtained using targeted liquid chromatography multiple reaction monitoring mass spectrometry, which is a highly specific, sensitive, and reproducible technique for quantifying targeted proteins. A list of protein fragments, or peptides, was sent to the detector. The samples then went through peak integration, outliers detection, normalization, quantification, and quality control using test/re-test samples. This procedure results in the intensity measures of 320 peptides that are annotated from 142 proteins. The brain imaging data were obtained using anatomical magnetic resonance imaging (MRI). Each image was first preprocessed following the standard pipeline, then mapped to an atlas consisting of 145 brain regions-of-interest to extract the volumetric measures \citep{doshi2016muse}. The atlas used in the study spans the entire brain and was actually built on multiple atlases. Individual atlases were first warped to the target image using a nonlinear registration method, followed by a spatially adaptive weighted voting strategy to fuse into a final segmentation. Moreover, the volume of each brain region was standardized by the total intracranial volume to account for variations of individual brain size. The cognitive outcome is a composite memory score, ADNI-MEM, that involves a battery of neuropsychological tests. In our study, we focus on 135 subjects diagnosed as mild cognitive impairment (MCI) patients at recruitment. MCI is a prodromal stage of AD, with a slight but noticeable and measurable decline in cognitive abilities. A person with MCI is at an increased risk of developing AD or other dementia. Understanding the pathologic mechanism underlying MCI provides important clues of onset of the disorder as well as a useful guide for early diagnosis and intervention.


\section{Model and Method}
\label{sec:model}

We first present the proposed model, then an estimation method integrating principal components analysis and penalized estimation.  

\subsection{Model}
\label{sub:model}

Suppose there are totally $n$ subjects. Let $\bX_{i} = (X_{i1},\ldots,X_{ir})^\top\in\mathbb{R}^{r}$ denote the $r$-dimensional vector of exposure variables, $\bM_{i} = (M_{i1},\ldots,M_{ip})^\top\in\mathbb{R}^{p}$ denote the $p$-dimensional vector of mediators, and $Y_{i}\in\mathbb{R}$ denote the univariate outcome variable, for subjects $i=1,\dots,n$. In our imaging proteomics study, $\bX_{i}$ denotes the protein peptide measures with $r=320$, $\bM_{i}$ denotes the brain volumetric measures with $p=145$, $Y_i$ denotes the memory score, and the sample size $n=135$. 

\begin{figure}[b!]
\begin{center}
\includegraphics[width=0.65\textwidth]{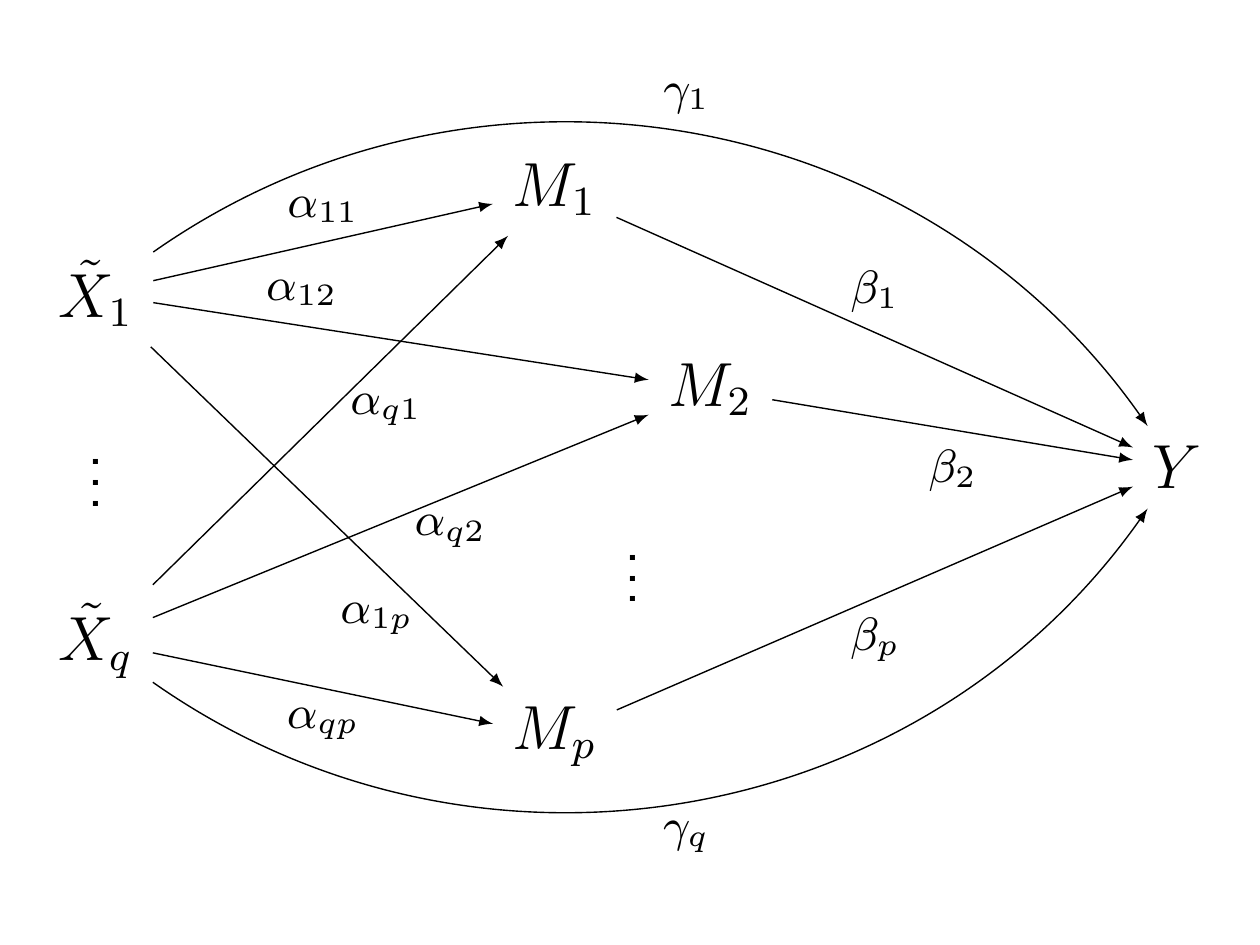}
\vspace{-0.3in}
\end{center}
\caption{\label{fig:dag}The schematic diagram of the proposed model with $q$ exposure variables $\tilde{X}_{1},\dots,\tilde{X}_{q}$, $p$ mediators $M_{1},\dots,M_{p}$, and the outcome variable $Y$.}
\end{figure}

The first step of our method is to perform a principal components analysis on $\bX_i$ to produce uncorrelated composite exposures. If $\bX_i$ further follows a multivariate normal distribution, then the produced composite exposures are independent. Let $\tilde{\bX}_{i} = (\tilde{X}_{i1},\ldots,\tilde{X}_{iq})^\top\in\mathbb{R}^{q}$ denote the first $q$ principal components. We then continue to model the path relations among $\tilde{\bX}_{i}, \bM_{i}$ and $Y_i$ via the following set of linear structural equation models, 
\begin{align}\label{eq:model}
\begin{split}
\bM &= \tilde{\bX}\balpha+\bepsilon, \\
\bY  &= \tilde{\bX}\bgamma+\bM\bbeta+\bdeta,
\end{split}
\end{align}
where $\tilde{\bX}=(\tilde{\bX}_{1},\dots,\tilde{\bX}_{n})^\top\in\mathbb{R}^{n\times q}$, $\bM=(\bM_{1},\dots,\bM_{n})^\top\in\mathbb{R}^{n\times p}$, $\bY=(Y_{1},\dots,Y_{n})^\top\in\mathbb{R}^{n}$ stack the composite exposures, mediators, and outcome across all subjects, respectively, $\bepsilon=(\bepsilon_{1},\dots,\bepsilon_{n})^\top\in\mathbb{R}^{n\times p}$, with $\bepsilon_{i}=(\epsilon_{i1},\dots,\epsilon_{ip})^\top\in\mathbb{R}^{p}$, and $\bdeta=(\eta_{1},\dots,\eta_{n})^\top\in\mathbb{R}^{n}$ are measurement errors. Suppose both error terms follow some zero mean normal distribution, and $\bepsilon$ is independent of $\tilde{\bX}$, $\bdeta$ is independent of $\tilde{\bX}$ and $\bM$, and $\bepsilon$ and $\bdeta$ are independent of each other. The parameters $\balpha=(\alpha_{jk})\in\mathbb{R}^{q\times p}$, $\bbeta=(\beta_{1},\dots,\beta_{p})^\top\in\mathbb{R}^{p}$, and $\bgamma=(\gamma_{1},\dots,\gamma_{q})^\top\in\mathbb{R}^{q}$ capture the path effects. Model \eqref{eq:model} is similar to that used in \citet{zhao2016pathway, zhao2020multimodal}, but none of those can handle multivariate exposure variables. Besides, we introduce some different forms of penalty functions in our parameter estimation. 

Figure~\ref{fig:dag} shows a schematic description of model \eqref{eq:model}. Under this model, we define the direct effect of $\tilde{X}_{j}$ on $Y$ as $\mathrm{DE}(\tilde{X}_{j})=\gamma_{j}$, the indirect effect of $\tilde{X}_{j}$ on $Y$ through $M_{k}$ as $\mathrm{IE}(\tilde{X}_{j},M_{k})=\alpha_{jk}\beta_{k}$, and the total indirect effect of $\tilde{X}_{j}$ on $Y$ as $\mathrm{IE}(\tilde{X}_{j})=\sum_{k=1}^{p}\alpha_{jk}\beta_{k}$, for $j=1,\dots,q$. The total effect of $\tilde{X}_{j}$ satisfies that $\mathrm{TE}(\tilde{X}_{j})=\mathrm{IE}(\tilde{X}_{j})+\mathrm{DE}(\tilde{X}_{j})=\sum_{k=1}^{p}\alpha_{jk}\beta_{k}+\gamma_{j}$. 

\begin{figure}[t!]
\begin{center}
\subfloat[\label{fig:dag-example-a}]{\includegraphics[width=0.45\textwidth]{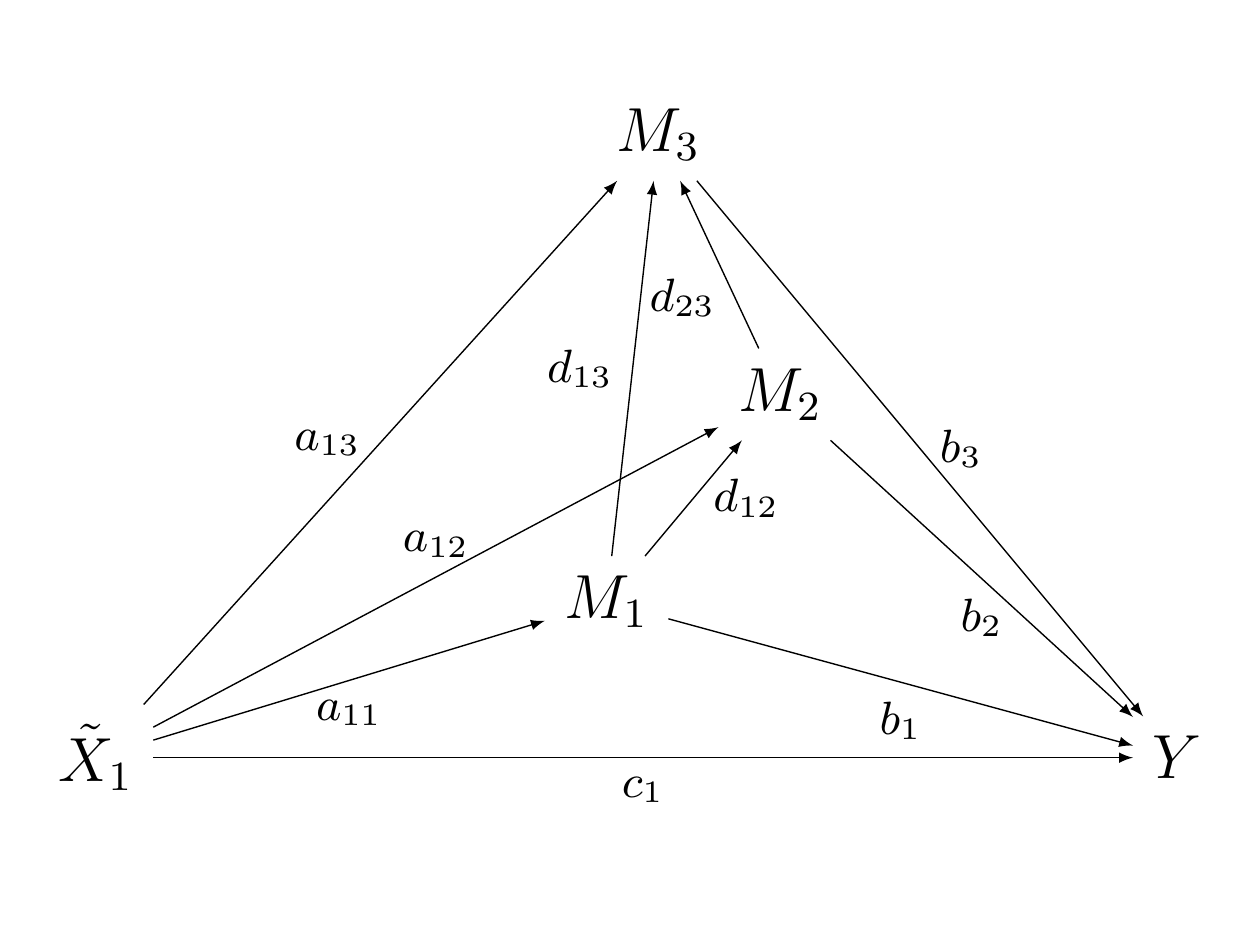}}
\enskip{}
\subfloat[\label{fig:dag-example-b}]{\includegraphics[width=0.45\textwidth]{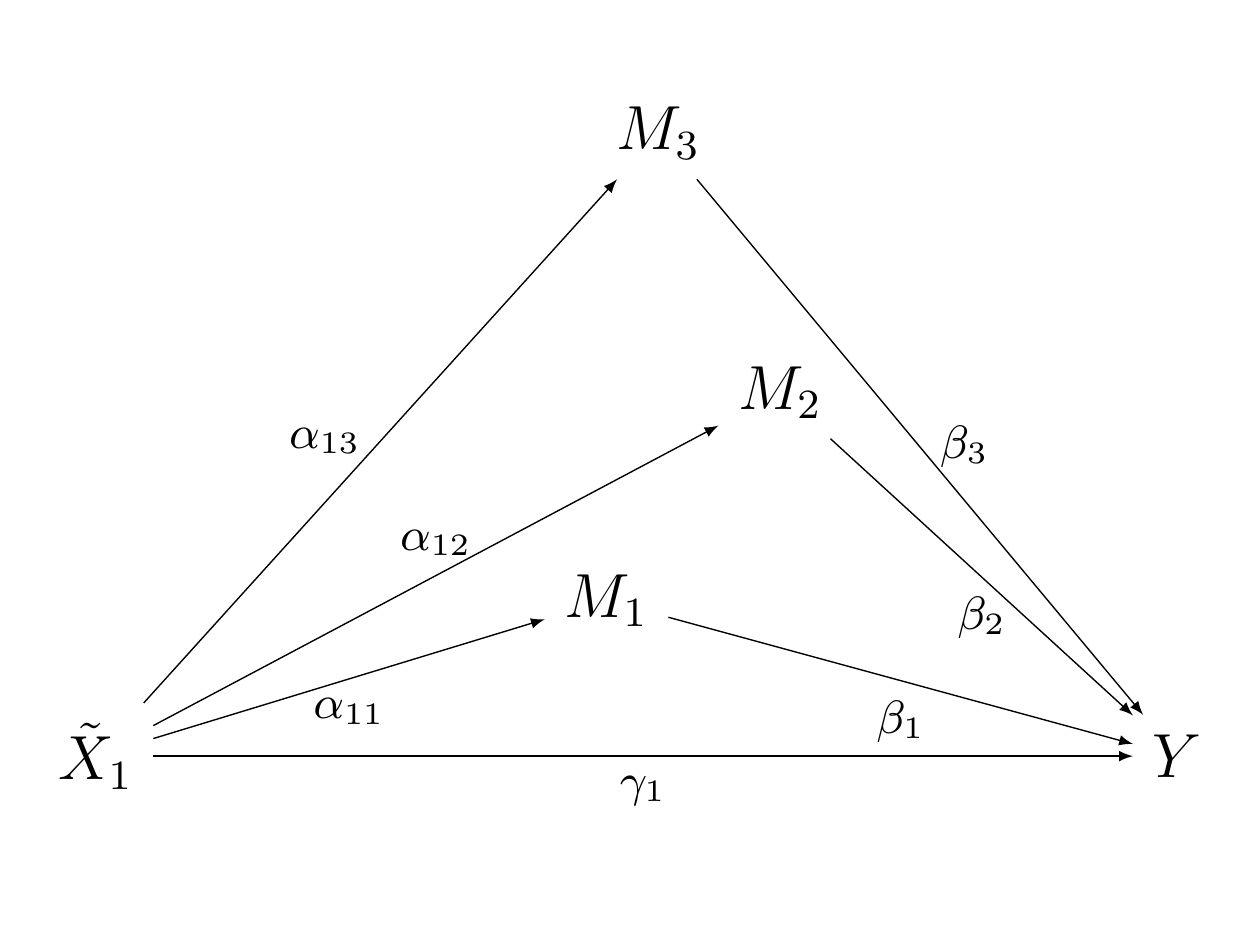}}
\vspace{-0.1in}
\end{center}
\caption{\label{fig:dag-example}A model example with $q=1$ exposure variable and $p=3$ sequentially ordered mediators.}
\end{figure}

A key characteristic of model \eqref{eq:model} is that it allows the multivariate mediators to be conditionally dependent given the exposures. To better illustrate this, we consider a simple example of model \eqref{eq:model}, where $q=1, p=3$, as shown in Figure~\ref{fig:dag-example}. In this example, Figure~\ref{fig:dag-example-a} outlines the sequential influences among all the mediators, while Figure~\ref{fig:dag-example-b} is the proposed model \eqref{eq:model}. We see that, for the first mediator, $M_1$, $\alpha_{11}=a_{11}, \beta_{1}=b_{1}$; for the second mediator, $M_{2}$, $\alpha_{12}=a_{11}d_{12}+a_{12}, \beta_{2}=b_{2}$; and for the third mediator, $M_3$, $\alpha_{13}=a_{11}d_{13} + a_{11}d_{12}d_{23} + a_{12}d_{23}+a_{13}, \beta_{3}=b_{3}$. As such, $\alpha_{1k}$ consolidates the effects through the $k$th mediator $M_k$, and the indirect effect $\mathrm{IE}(\tilde{X}_{1},M_{k})=\alpha_{1k}\beta_{k}$ can be viewed as the consolidated indirect effect through $M_{k}$, $k=1,2,3$.

\subsection{Estimation}
\label{sub:estimation}

We propose to estimate the parameters in model \eqref{eq:model} through the penalized ordinary least squares,
\begin{equation}\label{eq:opt_obj}
\underset{\balpha,\bbeta,\bgamma}{\text{minimize}} \;\; \frac{1}{2}\mathcal{L}(\balpha,\bbeta,\bgamma) + \lambda_{1}\mathcal{R}_{1}(\balpha,\bbeta) + \lambda_{2}\mathcal{R}_{2}(\balpha,\bbeta) + \lambda_{3}\mathcal{R}_{3}(\bgamma), 
\end{equation}
where the loss function is the usual least squares loss, 
\begin{equation*}
\mathcal{L}(\balpha,\bbeta,\bgamma) = \mathrm{tr}\left\{(\bM-\tilde{\bX}\balpha)^\top(\bM-\tilde{\bX}\balpha)\right\}+(\bY-\tilde{\bX}\bgamma-\bM\bbeta)^\top(\bY-\tilde{\bX}\bgamma-\bM\bbeta).
\end{equation*}
$\mathcal{R}_{1}, \mathcal{R}_{2}, \mathcal{R}_{3}$ are three penalty functions, with the tuning parameters $\lambda_1, \lambda_2, \lambda_3$, respectively. We next discuss each penalty function in detail. 

The first penalty function $\mathcal{R}_{1}$ is of the form,
\begin{equation*} \label{eq:pen_R1}
\mathcal{R}_{1}(\balpha,\bbeta) = \sum_{j=1}^{q}\sum_{k=1}^{p}\left\{|\alpha_{jk}\beta_{k}| + c_0\left( \alpha_{jk}^{2}+\beta_{k}^{2} \right)\right\} + c_1\left(\sum_{j=1}^{q}\sum_{k=1}^{p}|\alpha_{jk}|+\sum_{k=1}^{p}|\beta_{k}|\right),
\end{equation*}
for some parameters $c_0$ and $c_1$. It is a generalization of the pathway Lasso penalty of \citet{zhao2016pathway} to $q$ exposure variables, and is to facilitate selection of individual mediators. Specifically, for a given mediator $M_{k}$, the term $\sum_{j=1}^{q} |\alpha_{jk}\beta_{k}|$ is a product Lasso penalty, and encourages all the paths going through $M_k$ to be shrunk to zero, which in effect achieves the goal of mediator selection. The term $c_0( \alpha_{jk}^{2}+\beta_{k}^{2} )$ is to make the penalty a convex function, with a proper choice of the parameter $c_0$. It is straightforward to show that, when $c_0 \ge 1/2$, the sum $|\alpha_{jk}\beta_{k}| + c_0(\alpha_{jk}^{2}+\beta_{k}^{2})$ is convex. In our implementation, we fix $c_0 = 2$. The last term in $\mathcal{R}_{1}$ is the sum of usual Lasso penalty that further penalizes individual path effects $\alpha_{jk}, \beta_k$, with $c_1$ being an additional tuning parameter. It is found that this additional penalty helps further improves the selection accuracy \citep{zhao2016pathway}.

The second penalty function $\mathcal{R}_{2}$ is of the form,
\begin{equation*} \label{eq:pen_R2}
\mathcal{R}_{2}(\balpha,\bbeta) = \sum_{j=1}^{q}\sqrt{p}\sqrt{\sum_{k=1}^{p}(\alpha_{jk}\beta_{k})^{2}}.
\end{equation*}
It is a group Lasso penalty and is to facilitate the selection of individual exposure. Specifically, for a given exposure $\tilde{X}_{j}$, the penalty $\{\sum_{k=1}^{p}(\alpha_{jk}\beta_{k})^{2}\}^{1/2}$ encourages all the paths originating from $\tilde{X}_{j}$ to be shrunk to zero, which in effect achieves the goal of exposure selection.

The third penalty function $\mathcal{R}_{3}$ is of the form, 
\begin{equation*} \label{eq:pen_R3}
\mathcal{R}_{3}(\bgamma) = \sum_{j=1}^{q}|\gamma_{j}|.
\end{equation*}
This is simply the usual Lasso penalty and is to facilitate selection of direct effects between the exposures and the outcome. 

We next discuss how to solve the minimization problem \eqref{eq:opt_obj}. We note that \eqref{eq:opt_obj} involves the penalties on the product terms $\alpha_{jk}\beta_{k}$, making it difficult to derive the analytical solutions. As such, we first introduce a new parameter, $\mu_{jk}=\alpha_{jk}\beta_{k}$, which turns \eqref{eq:opt_obj} to an equivalent problem of solving a sparse group lasso that has an explicit form of solution \citep{simon2013sparse}. That is, letting $\bmu = (\mu_{jk}) \in \mathbb{R}^{q\times p}$, we turn to the equivalent optimization problem, 
\begin{align} \label{eq:opt_obj_const}
\begin{split}
\underset{\balpha,\bbeta,\bgamma,\bmu}{\text{minimize}} \;\;  \frac{1}{2} \mathcal{L}(\balpha,\bbeta,\bgamma) + \lambda_{1}\mathcal{R}_{1}(\bmu,\balpha,\bbeta) + \lambda_{2}\mathcal{R}_{2}(\bmu) + \lambda_{3}\mathcal{R}_{3}(\bgamma), \\
\text{ such that } \;\; \mu_{jk}=\alpha_{jk}\beta_{k}, \;\; \text{ for } j=1,\ldots,q \text{ and } k=1,\ldots,p.
\end{split}
\end{align}
Let $\bmu_{j}=(\mu_{j1},\dots,\mu_{jp})^\top\in\mathbb{R}^{p}$, $\balpha_{j}=(\alpha_{j1},\dots,\alpha_{jp})^\top\in\mathbb{R}^{p}$, and introduce the augmented Lagrangian parameter $\btau_{j}=(\tau_{j1},\dots,\tau_{jp})^\top\in\mathbb{R}^{p}$, for $j=1,\dots,q$, and $\btau = (\tau_{jk}) \in \mathbb{R}^{q\times p}$. Then, the augmented Lagrangian form of \eqref{eq:opt_obj_const} is,
\begin{align} \label{eq:opt_obj_lagrangian}
\begin{split}
\underset{\balpha,\bbeta,\bgamma,\bmu, \btau}{\text{minimize}} \;\; \frac{1}{2} \mathcal{L}(\balpha,\bbeta,\bgamma) + \lambda_{1}\mathcal{R}_{1}(\bmu,\balpha,\bbeta) + \lambda_{2}\mathcal{R}_{2}(\bmu) + \lambda_{3}\mathcal{R}_{3}(\bgamma) \\
+ \sum_{j=1}^{q}\left( \langle \bmu_{j}-\balpha_{j}\circ\bbeta,\btau_{j}\rangle+\frac{\rho}{2}\|\bmu_{j}-\balpha_{j}\circ\bbeta\|_{2}^{2} \right),
\end{split}
\end{align}
where $\rho>0$ is the augmented Lagrangian constant that we set $\rho=1$ in our implementation, $\circ$ is the Hadamard product, $\langle\cdot,\cdot\rangle$ is the inner product, and $\|\cdot\|_{2}$ is the $L_{2}$-norm. We next solve \eqref{eq:opt_obj_lagrangian} by updating $\bmu, \balpha, \bbeta, \bgamma$ and $\btau$ iteratively. 

More specifically, we first fix $\balpha^{(s)}, \bbeta^{(s)}, \bgamma^{(s)}, \btau^{(s)}$ at iteration $s$, and update $\bmu_{j}$ by solving
\begin{equation*} \label{eq:opt_obj_mu}
\underset{\bmu_{j}}{\text{minimize}}~\frac{\rho}{2}\|\bmu_{j}-\balpha_{j}^{(s)} \circ \bbeta^{(s)}\|_{2}^{2}+\btau_{j}^{(s)\top}\left( \bmu_{j}-\balpha_{j}^{(s)} \circ \bbeta^{(s)} \right) + \lambda_{1}\|\bmu_{j}\|_{1}+\lambda_{2}\sqrt{p}\|\bmu_{j}\|_{2},
\end{equation*}
for $j=1,\ldots,q$, where $\| \cdot \|_{1}$ is the $L_{1}$-norm. There is a closed-form solution,
\begin{equation}\label{eq:update_mu}
\begin{split}
\mu_{jk}^{(s+1)} = 
\begin{cases}
\left\{ \|\mathcal{S}(\bnu_{j},\lambda_{1}/\rho)\|_{2}-\lambda_{2}\sqrt{p}/\rho \right\}_{+}\frac{\mathcal{S}(\nu_{jk},\lambda_{1}/\rho)}{\|\mathcal{S}(\bnu_{j},\lambda_{1}/\rho)\|_{2}}, & \textrm{ if } \;\; \|\mathcal{S}(\bnu_{j},\lambda_{1}/\rho)\|_{2} \neq 0, \\
0 & \textrm{ otherwise},
\end{cases}
\end{split}
\end{equation}    
for $j=1,\ldots,q, k=1,\ldots,p$, where $\bnu_{j} = \balpha_{j}^{(s)}\circ\bbeta^{(s)}-\btau_{j}^{(s)}/\rho$, $\mathcal{S}(a, \lambda)=\mathrm{sgn}(a)\max\{|a|-\lambda,0\}$ is the soft-thresholding function with $\mathrm{sgn}(a)$ denoting the sign of $a$ and $a_{+}=\max\{a,0\}$, and $\mathcal{S}(\ba,\lambda)$ denotes the element-wise soft-thresholding of a vector $\ba$. 

We next fix $\bmu^{(s+1)}, \bbeta^{(s)}, \bgamma^{(s)}, \btau^{(s)}$, and update $\balpha_{j}$ by solving 
\begin{equation*} \label{eq:opt_alpha}
\underset{\balpha_{j}}{\text{minimize}}~\bV_{j}\balpha_{j}+\lambda_{1}c_{1}\mathrm{sgn}(\balpha_{j})-\bw_{j},  
\end{equation*}
where $\bV_{j}=\rho\bD_{\bbeta^{(s)}}^{2}+(4\lambda_{1}+\tilde{\bx}_{j}^\top\tilde{\bx}_{j})\boldsymbol{\mathrm{I}}_{p}$, $\bw_{j}=(\bM-\sum_{l\neq j}\tilde{\bx}_{l}\balpha_{l}^{(s)\top})^\top\tilde{\bx}_{j}+\bD_{\bbeta^{(s)}}\btau_{j}^{(s)}+\rho\bD_{\bbeta^{(s)}}\bmu_{j}^{(s+1)}$, $\bD_{\bbeta^{(s)}}$ is a diagonal matrix with $\bbeta^{(s)}$ as the diagonal elements, $\tilde{\bx}_{j}\in\mathbb{R}^{n}$ is the $j$th column of $\tilde{\bX}$, and $\boldsymbol{\mathrm{I}}_{p}$ is the $p$-dimensional identity matrix. The solution is 
\begin{equation}\label{eq:update_alpha}
\balpha_{j}^{(s+1)}=\bV_{j}^{-1}\mathcal{S}(\bw_{j},\lambda_{1}c_{1}), \quad j=1,\ldots,q.
\end{equation}

We next fix $\bmu^{(s+1)}, \balpha^{(s+1)}, \bgamma^{(s)}, \btau^{(s)}$, and update $\bbeta$ by solving 
\begin{equation*} \label{eq:opt_beta}
\underset{\bbeta}{\text{minimize}} ~\bV_{\bbeta}\bbeta+\lambda_{1}\mathrm{sgn}(\bbeta)-\bw_{\bbeta},
\end{equation*}
where $\bV_{\bbeta}=\bM^\top\bM+\rho\sum_{j=1}^{q}\bD_{\balpha_{j}^{(s+1)}}^{2}+4\lambda_{1} q\boldsymbol{\mathrm{I}}_{p}$, $\bw_{\bbeta}=\bM^\top(\bY-\tilde{\bX}\bgamma^{(s)}) + \sum_{j=1}^{q}\bD_{\balpha_{j}^{(s+1)}}\btau_{j}^{(s)}$ $+\rho\sum_{j=1}^{q}\bD_{\balpha_{j}^{(s+1)}}\bmu_{j}^{(s+1)}$, and $\bD_{\balpha_{j}^{(s+1)}}$ is a diagonal matrix with $\balpha_{j}^{(s+1)}$ as the diagonal elements. The solution is 
\begin{equation}\label{eq:update_beta}
\bbeta^{(s+1)}=\bV_{\bbeta}^{-1}\mathcal{S}(\bw_{\bbeta},\lambda_{1}c_{1}).
\end{equation}

We then fix $\bmu^{(s+1)}, \balpha^{(s+1)}, \bbeta^{(s+1)}, \btau^{(s)}$, and update $\bgamma$ by solving 
\begin{equation*} \label{eq:opt_gamma}
\underset{\bgamma}{\text{minimize}} ~\bV_{\bgamma}\bgamma+\lambda_{3}\mathrm{sgn}(\bgamma)-\bw_{\bgamma},
\end{equation*}
where $\bV_{\bgamma}=\tilde{\bX}^\top\tilde{\bX}$ and $\bw_{\bgamma}=\tilde{\bX}^\top(\bY-\bM\bbeta^{(s+1)})$. The solution is 
\begin{equation}\label{eq:update_gamma}
\bgamma^{(s+1)}=\bV_{\bgamma}^{-1}\mathcal{S}(\bw_{\bgamma},\lambda_{3}).
\end{equation}

Finally, we fix $\bmu^{(s+1)}, \balpha^{(s+1)}, \bbeta^{(s+1)}, \bgamma^{(s+1)}$, and update $\btau$ by
\begin{equation}\label{eq:update_tau}
\btau_{j}^{(s+1)}=\btau_{j}^{(s)}+\rho\left( \bmu_{j}^{(s+1)}-\balpha_{j}^{(s+1)}\circ\bbeta^{(s+1)} \right), \quad j=1,\ldots q.
\end{equation}

We stop the iterations until some stopping criterion is met. In our implement, we stop when the difference of two consecutive objective values is smaller than $10^{-6}$. We summarize the above optimization procedure in Algorithm~\ref{alg:opt_augLag}.

\begin{algorithm}[t!]
\caption{\label{alg:opt_augLag}The optimization algorithm for \eqref{eq:opt_obj_lagrangian}.}
\begin{algorithmic}[1]
\INPUT $(\tilde{\bX},\bM,\bY)$ and the tuning parameters $\lambda_1, \lambda_2, \lambda_3$ and $c_{1}$.
\State \textbf{initialization}: $\left\{\balpha^{(0)},\bbeta^{(0)},\bgamma^{(0)},\bmu^{(0)},\btau^{(0)}\right\}$.
\Repeat    
\State update $\mu_{jk}^{(s+1)}$ given $\balpha^{(s)}, \bbeta^{(s)}, \bgamma^{(s)}, \btau^{(s)}$ by \eqref{eq:update_mu}, for $j=1,\ldots,q, k=1,\ldots,p$. 
\State update $\balpha_{j}^{(s+1)}$ given $\bmu^{(s+1)}, \bbeta^{(s)}, \bgamma^{(s)}, \btau^{(s)}$ by \eqref{eq:update_alpha}, for $j=1,\ldots,q$.
\State update $\bbeta^{(s+1)}$ given $\bmu^{(s+1)}, \balpha^{(s+1)}, \bgamma^{(s)}, \btau^{(s)}$ by \eqref{eq:update_beta}.
\State update $\bgamma^{(s+1)}$ given $\bmu^{(s+1)}, \balpha^{(s+1)}, \bbeta^{(s+1)}, \btau^{(s)}$ by \eqref{eq:update_gamma}.
\State update $\btau_j^{(s+1)}$ given $\bmu^{(s+1)}, \balpha^{(s+1)}, \bbeta^{(s+1)}, \bgamma^{(s+1)}$ by \eqref{eq:update_tau}, for $j=1,\ldots,q$. 
\Until{the stopping criterion is met.}
\OUTPUT $\left\{\hat{\balpha},\hat{\bbeta},\hat{\bgamma}\right\}$.
\end{algorithmic}
\end{algorithm}

We tune the parameters in \eqref{eq:opt_obj_lagrangian} using the Bayesian information criterion (BIC),
\begin{equation*} \label{eq:BIC}
\mathrm{BIC}=-2\log{\mathcal{L}\left(\hat{\balpha},\hat{\bbeta},\hat{\bgamma}\right)}+\log(n)|\hat{\mathcal{A}}|,
\end{equation*}
where $\hat{\balpha},\hat{\bbeta},\hat{\bgamma}$ are the estimates under a given set of tuning parameters $\lambda_1, \lambda_2, \lambda_3$ and $c_{1}$, $\mathcal{A}=\left\{(j,k):\alpha_{jk}\beta_{k}\neq 0\right\}$ denotes the active set, and $|\mathcal{A}|$ is the cardinality. In our implementation, we adopt the tuning strategy of \citet{zou2005regularization}, by tuning the ratios $\lambda_{2}/\lambda_{1}, \lambda_{3}/\lambda_{1}$ along with $c_{1}$ in a grid search, and choose the best set of parameters that minimizes the BIC.


\section{AD Imaging Proteomics Study Revisited}
\label{sec:ADNI}

We apply the proposed method to the ADNI imaging proteomics data, taking the CSF peptide measures as the exposures, the brain volumetric measures as the mediators, and the memory score as the outcome. Moreover, we adjust the exposures, mediators, and outcome for age, gender, ApoE4, and years of education to remove potential confounding effects \citep{rosenbaum2002covariance}. We first summarize the identified paths with nonzero effects, then discuss the relevant proteins and brain regions in detail. In summary, our findings are consistent with the existing knowledge of AD. Moreover, our method also suggests a few potentially interesting protein-structure-memory paths that may deserve further examination and verification.

\subsection{Paths with nonzero effects}
\label{sub:pathways}

\begin{figure}
\begin{center}
\includegraphics[width=0.875\textwidth,angle=-90]{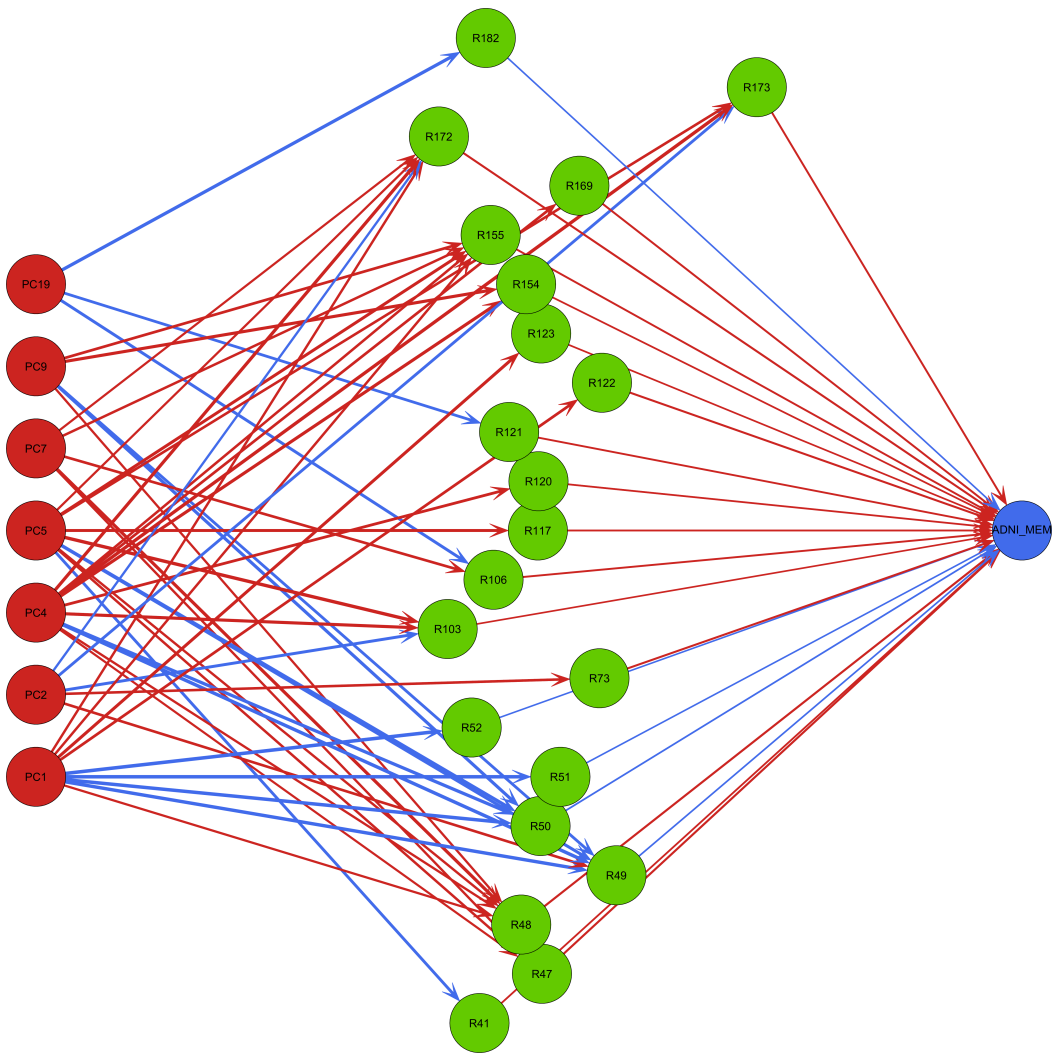}
\vspace{-0.1in}
\end{center}
\caption{\label{fig:ADNI_path} The estimated paths for the AD imaging proteomics study. The red nodes denote the principal components of the peptides as exposures, the green nodes the brain regions as mediators, and the blue node the memory score as outcome. The red arrows indicate positive path effects, and the blue arrows negative path effects.}
\end{figure}

We first apply principal components analysis to the peptide data. The top 20 principal components (PCs) account for about $85\%$ of total data variation. We thus focus on those $q=20$ top PCs and feed them as the exposure variables into the subsequent penalized path analysis. Figure~\ref{fig:ADNI_path} presents all the identified paths with a nonzero indirect path effect. Table~\ref{tab:ADNI_IE} presents the estimated path effects including the estimated $\alpha$ and $\beta$ of each path, and Table~\ref{tab:ADNI_PCeffect} presents the indirect, direct, and total effect of each exposure PC. 

\begin{table}
\caption{\label{tab:ADNI_IE} The brain regions with a nonzero indirect effect ($\mathrm{IE}=\alpha\beta$) for the AD imaging proteomics study.}
\begin{center}
\vskip-20pt
\resizebox{\textwidth}{!}{
{\small
\begin{tabular}{l l l r r r r r r r r}
      \hline
      & & & \multicolumn{7}{c}{Principal components of peptides as exposures} \\
      \cline{4-10}
      & \multicolumn{1}{c}{\multirow{-2}{*}{Brain regions as mediators}} & & \multicolumn{1}{c}{PC1} & \multicolumn{1}{c}{PC2} & \multicolumn{1}{c}{PC4} & \multicolumn{1}{c}{PC5} & \multicolumn{1}{c}{PC7} & \multicolumn{1}{c}{PC9} & \multicolumn{1}{c}{PC19} & \multicolumn{1}{c}{\multirow{-2}{*}{$\beta$ ($\times 10^{-2}$)}} \\
      \hline
       & & $\alpha$ & - & - & - & -0.17 & - & - & - & \\
      \multirow{-2}{*}{R41} & \multirow{-2}{*}{Left cerebellum white matter} & IE ($\times 10^{-3}$) & - & - & - & -1.30 & - & - & - & \multirow{-2}{*}{0.76} \\ 
      \hline
      & & $\alpha$ & - & - & 0.11 & 0.13 & 0.13 & - & - \\
      \multirow{-2}{*}{R47} & \multirow{-2}{*}{Right hippocampus} & IE ($\times 10^{-3}$) & - & - & 1.12 & 1.60 & 1.52 & - & - & \multirow{-2}{*}{1.17} \\
      \hline
      & & $\alpha$ & 0.11 & - & 0.13 & 0.13 & 0.22 & 0.12 & - \\
      \multirow{-2}{*}{R48} & \multirow{-2}{*}{Left hippocampus} & IE ($\times 10^{-3}$) & 1.34 & - & 1.59 & 1.76 & 3.40 & 1.55 & - & \multirow{-2}{*}{1.20} \\
      \hline
      & & $\alpha$ & -0.25 & 0.15 & -0.26 & -0.18 & - & -0.16 & - \\
      \multirow{-2}{*}{R49} & \multirow{-2}{*}{Temporal horn of right lateral ventricle} & IE ($\times 10^{-3}$) & 2.06 & -1.01 & 2.03 & 1.28 & - & 1.08 & - & \multirow{-2}{*}{-0.66} \\
      \hline
      & & $\alpha$ & -0.29 & - & -0.23 & -0.25 & - & -0.21 & - \\
      \multirow{-2}{*}{R50} & \multirow{-2}{*}{Temporal horn of left lateral ventricle} & IE ($\times 10^{-3}$) & 2.55 & - & 1.78 & 2.05 & - & 1.74 & - & \multirow{-2}{*}{-0.71} \\
      \hline
      & & $\alpha$ & -0.36 & - & - & - & - & - & - \\
      \multirow{-2}{*}{R51} & \multirow{-2}{*}{Right lateral ventricle} & IE ($\times 10^{-3}$) & 1.06 & - & - & - & - & - & - & \multirow{-2}{*}{-0.30} \\
      \hline
      & & $\alpha$ & -0.36 & - & - & - & - & - & - \\
      \multirow{-2}{*}{R52} & \multirow{-2}{*}{Left lateral ventricle} & IE ($\times 10^{-3}$) & 1.15 & - & - & - & - & - & - & \multirow{-2}{*}{-0.27} \\
      \hline
      & & $\alpha$ & - & 0.14 & - & - & - & - & - \\
      \multirow{-2}{*}{R73} & \multirow{-2}{*}{Cerebellar vermal lobules VIII-X} & IE ($\times 10^{-3}$) & - & 1.88 & - & - & - & - & - & \multirow{-2}{*}{1.00} \\
      \hline
      & & $\alpha$ & - & -0.17 & 0.20 & 0.25 & - & - & - \\
      \multirow{-2}{*}{R103} & \multirow{-2}{*}{Left anterior insula} & IE ($\times 10^{-3}$) & - & -1.13 & 1.27 & 1.76 & - & - & - & \multirow{-2}{*}{0.56} \\
      \hline
      & & $\alpha$ & - & - & - & - & 0.15 & - & -0.18 \\
      \multirow{-2}{*}{R106} & \multirow{-2}{*}{Right angular gyrus} & IE ($\times 10^{-3}$) & - & - & - & - & 1.03 & - & -1.41 & \multirow{-2}{*}{0.78} \\
      \hline
      & & $\alpha$ & - & - & - & 0.17 & - & - & - \\
      \multirow{-2}{*}{R117} & \multirow{-2}{*}{Left entorhinal areas} & IE ($\times 10^{-3}$) & - & - & - & 1.15 & - & - & - & \multirow{-2}{*}{0.76} \\
      \hline
      & & $\alpha$ & - & - & 0.16 & - & - & - & - \\
      \multirow{-2}{*}{R120} & \multirow{-2}{*}{Right frontal pole} & IE ($\times 10^{-3}$) & - & - & 1.12 & - & - & - & - & \multirow{-2}{*}{0.75} \\
      \hline
      & & $\alpha$ & - & - & - & - & - & - & -0.16 \\
      \multirow{-2}{*}{R121} & \multirow{-2}{*}{Left frontal pole} & IE ($\times 10^{-3}$) & - & - & - & - & - & - & -1.12 & \multirow{-2}{*}{0.77} \\
      \hline
      & & $\alpha$ & 0.16 & - & - & - & - & - & - \\
      \multirow{-2}{*}{R122} & \multirow{-2}{*}{Right fusiform gyrus} & IE ($\times 10^{-3}$) & 1.32 & - & - & - & - & - & - & \multirow{-2}{*}{1.02} \\
      \hline
      & & $\alpha$ & 0.19 & - & - & - & - & - & - \\
      \multirow{-2}{*}{R123} & \multirow{-2}{*}{Left fusiform gyrus} & IE ($\times 10^{-3}$) & 1.12 & - & - & - & - & - & - & \multirow{-2}{*}{0.66} \\
      \hline
      & & $\alpha$ & - & - & 0.21 & - & - & 0.19 & - \\
      \multirow{-2}{*}{R154} & \multirow{-2}{*}{Right middle temporal gyrus} & IE ($\times 10^{-3}$) & - & - & 1.68 & - & - & 1.66 & - & \multirow{-2}{*}{0.74} \\
      \hline
      & & $\alpha$ & 0.13 & - & 0.14 & 0.18 & 0.14 & 0.15 & - \\
      \multirow{-2}{*}{R155} & \multirow{-2}{*}{Left middle temporal gyrus} & IE ($\times 10^{-3}$) & 1.01 & - & 1.09 & 1.63 & 1.05 & 1.35 & - & \multirow{-2}{*}{0.79} \\
      \hline
      & & $\alpha$ & - & - & 0.15 & - & - & - & - \\
      \multirow{-2}{*}{R169} & \multirow{-2}{*}{Left precuneus} & IE ($\times 10^{-3}$) & - & - & 1.10 & - & - & - & - & \multirow{-2}{*}{0.82} \\
      \hline
      & & $\alpha$ & 0.13 & -0.12 & 0.22 & 0.11 & 0.11 & - & - \\
      \multirow{-2}{*}{R172} & \multirow{-2}{*}{Right posterior insula} & IE ($\times 10^{-3}$) & 1.44 & -1.30 & 2.67 & 1.03 & 1.00 & - & - & \multirow{-2}{*}{1.03} \\
      \hline
      & & $\alpha$ & - & -0.17 & 0.24 & 0.15 & - & - & - \\
      \multirow{-2}{*}{R173} & \multirow{-2}{*}{Left posterior insula} & IE ($\times 10^{-3}$) & - & -1.46 & 2.18 & 1.09 & - & - & - & \multirow{-2}{*}{0.82} \\
      \hline
      & & $\alpha$ & - & - & - & - & - & - & -0.24 \\
      \multirow{-2}{*}{R182} & \multirow{-2}{*}{Right precentral gyrus} & IE ($\times 10^{-3}$) & - & - & - & - & - & - & 1.91 & \multirow{-2}{*}{-0.51} \\
      \hline
\end{tabular}
}}
\end{center}
\end{table}

\begin{table}
\caption{\label{tab:ADNI_PCeffect} The estimated indirect effects (IE), direct effects (DE), and total effects (TE) of the top principal components. The PCs with zero IE and DE are not presented in the table.}
\begin{center}
\resizebox{\textwidth}{!}{
\begin{tabular}{l r r r r r r r r r r r r r}
\hline
& \multicolumn{1}{c}{PC1} & \multicolumn{1}{c}{PC2} & \multicolumn{1}{c}{PC4} & \multicolumn{1}{c}{PC5} & \multicolumn{1}{c}{PC6} & \multicolumn{1}{c}{PC7} & \multicolumn{1}{c}{PC9} & \multicolumn{1}{c}{PC11} & \multicolumn{1}{c}{PC14} & \multicolumn{1}{c}{PC15} & \multicolumn{1}{c}{PC16} & \multicolumn{1}{c}{PC19} & \multicolumn{1}{c}{Total} \\
\hline
IE & 0.013 & -0.003 & 0.018 & 0.012 & - & 0.008 & 0.007 & - & - & - & - & -0.001 & 0.054 \\
DE & 0.138 & - & - & 0.066 & -0.035 & 0.168 & 0.065 & -0.018 & 0.102 & -0.007 & 0.156 & - & 0.634 \\
TE & 0.151 & -0.003 & 0.018 & 0.078 & -0.035 & 0.176 & 0.072 & -0.018 & 0.102 & -0.007 & 0.156 & -0.001 & 0.688 \\
\hline
\end{tabular}
}
\end{center}
\end{table}

\subsection{Proteins}
\label{sub:proteins}

Among the 20 PCs, seven have nonzero indirect effects on memory. Next, we focus on PC1, PC4, and PC5 as they account for a higher proportion of total data variation and demonstrate a relatively higher indirect path effect on the outcome. Table \ref{tab:ADNI_protein} lists the top proteins in PC1, PC4, and PC5, and the corresponding gene name. We also include the regulation directions found in the AD literature, where an upregulation compared to cognitive normal controls indicates a higher protein abundance in MCI/AD patients, as well as the direction of correlations with the CSF amyloid-$\beta$ and tau, the two well-established AD protein biomarkers \citep{wesenhagen2020cerebrospinal}. We next discuss the identified proteins by their relevance in the amyloid-$\beta$ and tau pathology.

\begin{table}
\caption{\label{tab:ADNI_protein}Proteins with top loading magnitude in PC1, PC4 and PC5. For each protein, direction of protein level in MCI/AD compared to normal control and correlation with CSF tau and amyloid reported in the literature are provided.}
\begin{center}
\resizebox{\textwidth}{!}{
{\small
\begin{tabular}{L{8cm} r l c c c}
      \hline
      & & & & \multicolumn{2}{c}{Correlation} \\
      \cline{5-6}
      \multicolumn{1}{c}{\multirow{-2}{*}{Protein}} & \multicolumn{1}{c}{\multirow{-2}{*}{Loading}} & \multicolumn{1}{c}{\multirow{-2}{*}{Gene}} & \multicolumn{1}{c}{\multirow{-2}{*}{Direction}} & \multicolumn{1}{c}{tau} & \multicolumn{1}{c}{amyloid} \\
      \hline
      \textbf{PC1} \\
      ProSAAS & 0.075 & PCSK1N & $\Updownarrow$ & \\
      Neuronal growth regulator 1 & 0.075 & NEGR1 & $\downarrow$ \\
      Cell adhesion molecule 3 & 0.075 & CADM3 & $\downarrow$ \\
      Neuroblastoma suppressor of tumorigenicity 1 & 0.073 & NBL1 & $\uparrow$ \\
      Spondin-1 & 0.073 & SPON1 & $\uparrow$ & $\uparrow$ & $\downarrow$ \\
      Prostagiandin-H2 D-isomerase & 0.073 & PTGDS & $\downarrow$ & & $\downarrow$ \\
      Monocyte differentiation antigen CD14 & 0.071 & CD14 & $\uparrow$ \\
      VPS10 domain-containing receptor SorCS1 & 0.069 & SORCS1 & & $\uparrow$ & $\downarrow$ \\
      \textbf{PC4} \\
      Neuronal pentraxin-2 & 0.152 & NPTX2 & $\downarrow$ & & $\uparrow$ \\
      Insulin-like growth factor-binding protein 2 & -0.146 & IGFBP2 & $\Updownarrow$ & $\uparrow$ \\
      Beta-2-microglobulin & -0.125 & B2M & $\Updownarrow$ & $\downarrow$ \\
      Neurexin-2 & 0.116 & NRXN2 & $\Updownarrow$ \\
      Apolipoprotein D & -0.095 & APOD & $\Updownarrow$ & $\uparrow$ \\
      Neuronal pentraxin-1 & 0.093 & NPTX1 & $\downarrow$ \\
      Kallikrein-6 & -0.083 & KLK6 & $\uparrow$ & $\uparrow$ & $\uparrow$ \\
      Cystatin-C & -0.066 & CST3 & $\Updownarrow$ & $\Updownarrow$ & $\uparrow$ \\
      \textbf{PC5} \\
      Complement C4-A & -0.180 & C4A & $\uparrow$ \\
      Ectonucleotide pyrophosphatase/phosphodiesterase family member 2 & -0.144 & ENPP2 & $\uparrow$ & $\downarrow$ \\
      Superoxide dismutase [Cu-Zn] & 0.129 & SOD1 & $\downarrow$ & $\uparrow$ & $\downarrow$ \\
      Complement factor B & 0.110 & CFB & $\uparrow$ \\
      Glial fibrillary acidic protein & -0.106 & GFAP & $\uparrow$\\
      Chromogranin-A & 0.105 & CHGA & $\Updownarrow$ & $\uparrow$ & $\uparrow$ \\
      Mimecan & -0.094 & OGN & $\Updownarrow$ \\
      Neurosecretory protein VGF & 0.083 & VGF & $\downarrow$ & & $\downarrow$ \\
      Alpha-1B-glycoprotein & 0.075 & A1BG & $\Updownarrow$ \\
      \hline
\end{tabular}
}}
\end{center}
\vspace{-1ex}
{\raggedright {\scriptsize{$\uparrow$: consistently upregulated in MCI/AD or positively correlated; $\downarrow$: consistently downregulated in MCI/AD or negatively correlated; $\Updownarrow$: inconsistent reports.}}}
\end{table}

\underline{\textit{Proteins related to amyloid pathology.}} Among the top-loaded proteins, SPON1, SORCS1, PTGDS, CST3, NPTX2, VGF, and CHGA have been found to be related to  amyloid-$\beta$ pathology in AD. The accumulation of amyloid-$\beta$ is generally considered a hallmark of AD, which is derived from the amyloid precursor protein (APP) through sequential cleavages by beta-site amyloid precursor protein cleaving enzyme 1 (BACE1) and $\gamma$-secretase~\citep{vassar1999beta}. Blocking BACE1 can potentially reduce the abundance in amyloid-$\beta$, however, this may prohibit the other functions of BACE1 in psychological activities. 
For SPON1, using an in vivo AD mouse model, it was found that, by injecting SPON1, the amount of amyloid-$\beta$ was significantly reduced, and subsequently, the ameliorated cognitive dysfunction and memory impairment were improved, suggesting SPON1 to be a potential AD therapy target \citep{park2020spon1}. Interacting with APOE, human SPON1 suppresses amyloid-$\beta$ level through the APP transgene, and has an impact on working memory performance through the activation of the triangular part of the right inferior frontal gyrus~\citep{liu2018apoe}. 
For NPTX1 and NPTX2, both belong to the family of long neuronal pentraxins. Together with NPTXR, they bind AMPA type glutamate receptors and contribute to multiple forms of developmental and adult synaptic plasticity. Using an AD mouse model, reduction in NPTX2 together with amyloidosis was found to induce a synergistic reduction of inhibitory circuit function. In AD subjects, the level of NTPX2 was found to be related to hippocampal volume, as well as cognitive decline \citep{xiao2017nptx2}. 
For CST3, cysteine proteases, including cathepsin B (CatB), is a recently discovered amyloid-$\beta$-degrading enzyme. Using a mouse model, CST3 was discovered to be a key inhibitor of CatB-induced amyloid-$\beta$ degradation in vivo. Genetic ablation of CST3 significantly reduced soluble amyloid-$\beta$ levels, and attenuated associated cognitive deficits and behavioral abnormalities, and restored synaptic plasticity in hippocampus \citep{sun2008cystatin}. 
For VGF, through a mouse model, over-expression of neuropeptides precursor VGF was found to partially rescue amyloid-$\beta$-mediated memory impairment and neuropathology, suggesting a possible causal role of VGF in protecting against AD pathogenesis and progression \citep{beckmann2020multiscale}.
For SORCS1, through a meta-analysis of 16 SORCS1-single nucleotide polymorphisms (SNPs) in six independent datasets, it was found that over-expression of SORCS1 can reduce $\gamma$-secretase activity and amyloid-$\beta$ levels, and the suppression of SORCS1 can increase $\gamma$-secretase processing of APP and the levels of amyloid-$\beta$ \citep{reitz2011sorcs1}. 
For PTGDS, it is one of the most abundant proteins in the CSF, which binds and transports small lipophilic molecules such as amyloid-$\beta$, and thus has been considered as the endogenous amyloid-$\beta$ chaperone \citep{kanekiyo2007lipocalin}, and is believed to play an important role in AD development. 
For CHGA, compared to the normal controls, the level of CHGA was significantly higher in the CSF of patients with MCI, especially with MCI progressing to AD \citep{duits2018synaptic}. CHGA is the major soluble protein in catecholamine storage vesicles, abnormalities of which may play a central role in memory deficits in AD. Elevation of CHGA was observed in AD brains, and was believed to play a role in amyloid-$\beta$ pathology \citep{o1993cerebrospinal,mattsson2013csf}. It has also been found that CHGA is negatively associated with hippocampal and entorhinal volume~\citep{khan2015subset}.

\underline{\textit{Proteins related to tau pathology.}}
For IGFBP2, it is an abundant cerebral insulin-like growth factor signaling protein associated with the AD biomarkers. In both AD mouse models and AD patients, IGFBP2 was observed to be associated with CSF tau levels and brain atrophy in non-hippocampal regions, suggesting that it is relevant in neurodegeneration through tau pathology \citep{bonham2018insulin}.

\underline{\textit{Proteins related to both amyloid and tau pathology.}}
There was evidence showing that proteins KLK6 and SOD1 were relevant in both amyloid and tau pathology.
For SOD1, using an APP-overexpressing mouse model, SOD1 deficiency was found to accelerate amyloid-$\beta$ oligomerization, induce tau phosphorylation and lower levels of synaptophysin, and consequently memory impairment \citep{murakami2011sod1}. 
Kallikrein-related peptidases (KLKs) represent the largest family of secreted serine proteases. Human KLK6 is the most abundant KLKs in the spinal cord, brain stem, cerebral cortex including the hippocampus and thalamus. It has been found that KLK6 cleaves APP and mediates cleavage of laminin and collagen, which has implications for APP processing and amyloid-$\beta$ mediated neurotoxicity \citep{small1993role,angelo2006substrate}. In AD patients, the level of KLK6 in CSF is significantly elevated and is associated with levels of CSF tau suggesting a potential marker of tau pathology \citep{goldhardt2019kallikrein}.

\underline{\textit{Other AD related proteins.}}
NRXN2 is another protein marker that was found to be up-regulated among MCI patients, especially with MCI progression to AD \citep{duits2018synaptic}.
APOD was found to be elevated in the prefrontal cortex associated with cognitive decline \citep{thomas2003apolipoprotein}. 
GFAP immunohistochemistry is a marker to assess the oxidative stress and glial cell activation expressed in astrocytes. Focusing on the human entorhinal cortex and hippocampus, the GFAP expression was observed in the hippocampus of AD patients \citep{hol2003neuronal}. 
B2M is a component of major histocompatibility complex class 1 molecules. Increased soluble B2M has been discovered in the CSF of patients with AD, and was associated with cognitive decline \citep{carrette2003panel}. Using mouse models, elevated B2M was observed in the hippocampus of aged mice. Injecting exogenous B2M locally in the hippocampus, impaired hippocampal-dependent cognitive function and neurogenesis were observed in young mice. The findings suggest that the accumulation of B2M increases the risk of age-related cognitive dysfunction and neurogenesis impairment \citep{smith2015beta2}.

\underline{\textit{Proteins related to brain structure/atrophy.}}
NEGR1 is a member of the immunoglobulin superfamily of cell adhesion molecules, and is involved in cortical layering. Using a NEGR1-targeted mouse model, brain morphological analysis revealed NEGR1-related neuroanatomical abnormalities, including enlargement of ventricles and decrease in the volume of the whole brain, corpus callosum, globus pallidus, and hippocampus \citep{singh2019neural}. 
CST3 was discovered to be related to a higher hippocampal atrophy rate \citep{paterson2014cerebrospinal}, and atrophy in the entorhinal cortex \citep{mattsson2014effects}. 
APOD and NPTX2 were found to be related to medial temporal lobe atrophy \citep{mattsson2014effects,swanson2016neuronal}.

\subsection{Brain regions}
\label{sub:regions}

\begin{figure}
\begin{center}
\subfloat[PC1]{\includegraphics[width=0.33\textwidth]{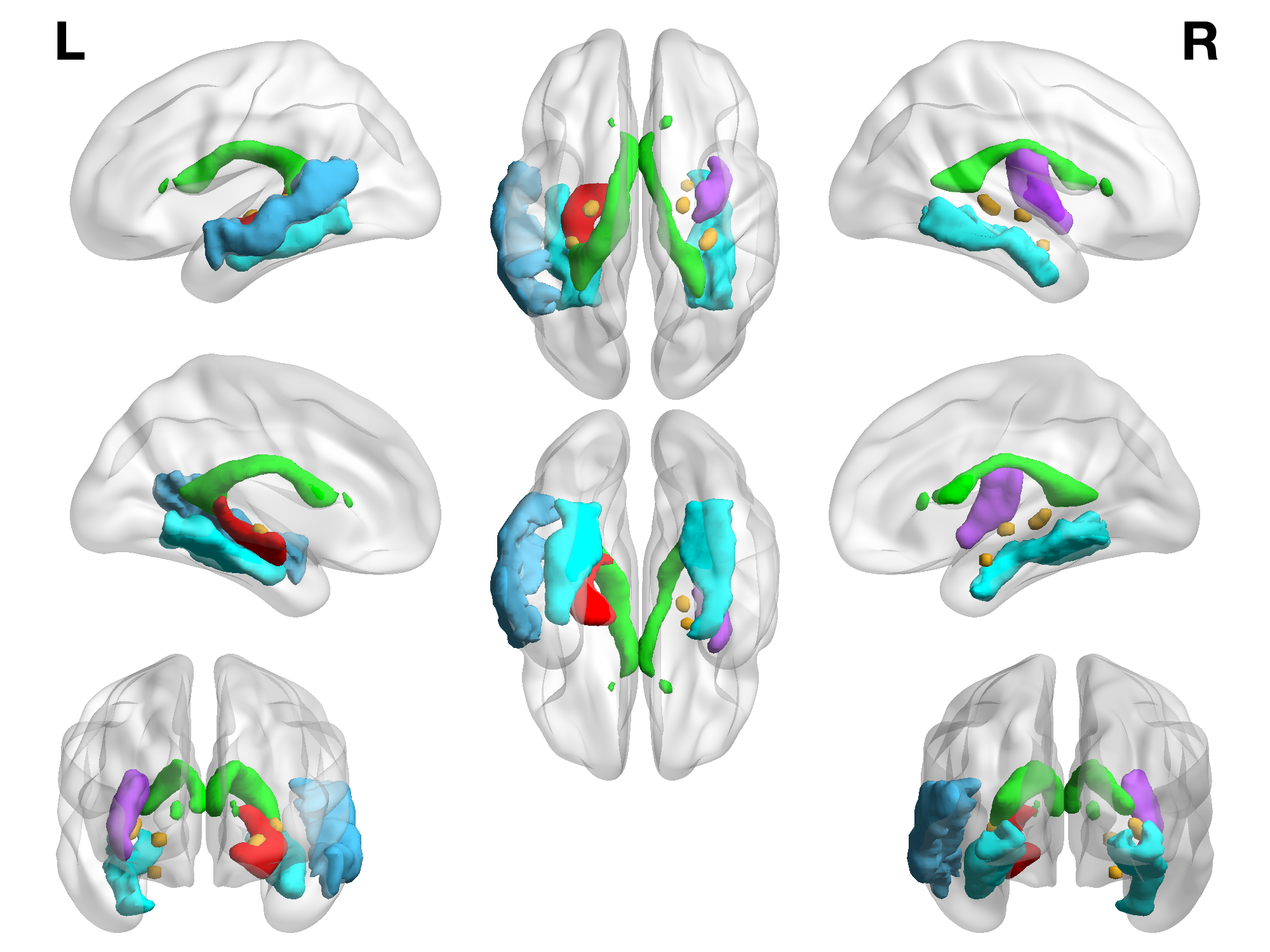}}
\subfloat[PC4]{\includegraphics[width=0.33\textwidth]{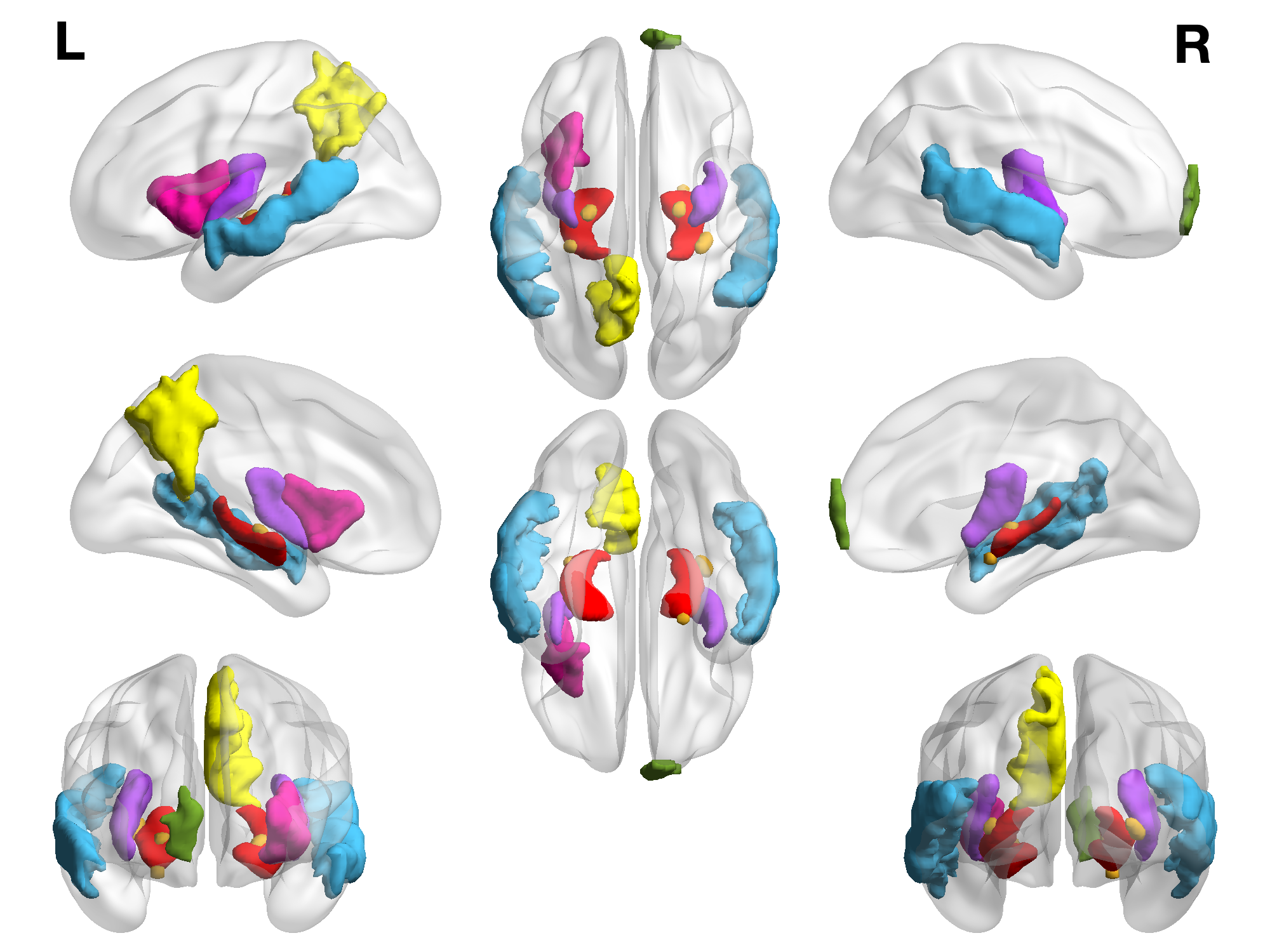}}
\subfloat[PC5]{\includegraphics[width=0.33\textwidth]{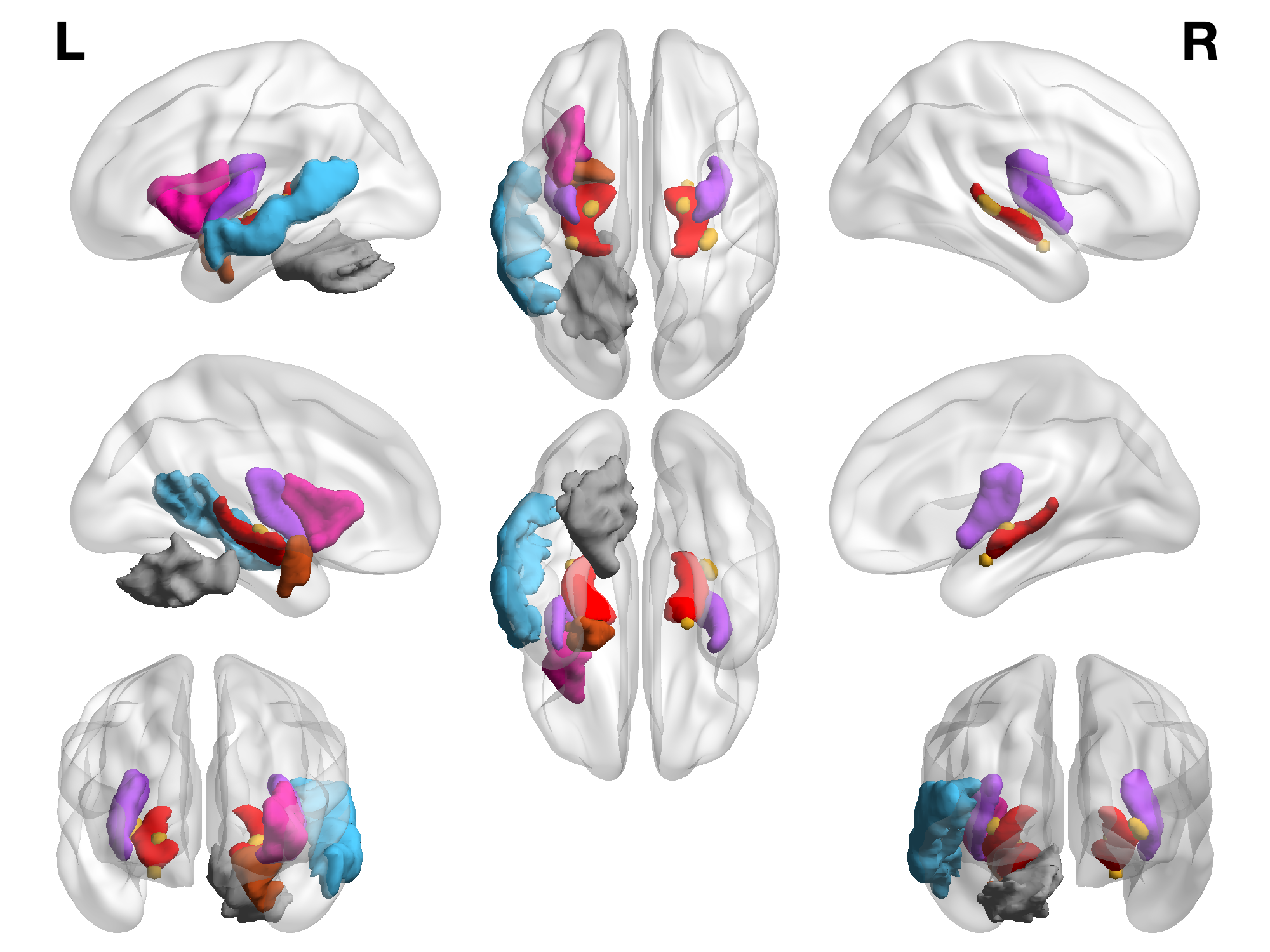}}
\vspace{1ex}
\includegraphics[width=\textwidth,trim=0cm 5cm 20cm 5cm,clip]{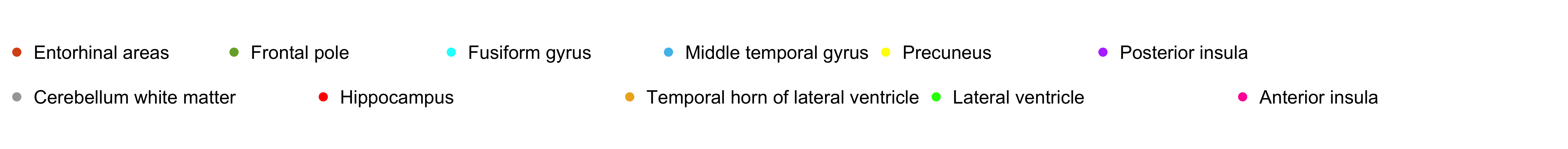}
\vspace{-0.1in}
\end{center}
\caption{\label{fig:ADNI_PCROI}Brain regions with a nonzero mediation effect in (a) PC1, (b) PC4, and (c) PC5.}
\end{figure}

While Table \ref{tab:ADNI_IE} lists the brain regions with nonzero path effects induced by PC1, PC4, and PC5, Figure~\ref{fig:ADNI_PCROI} visualizes those regions on a template brain. The identified brain regions include the hippocampus, the entorhinal cortex, cortical regions on the temporal, parietal and frontal lobes, the lateral ventricles, and the cerebellum. Brain structural atrophy occurs early in the medial temporal lobe, including the hippocampus and entorhinal cortex, then extends soon after to the rest of the cortical areas, usually following a temporal, parietal, frontal trajectory, whereas the motor areas are affected toward late stages. \citep{pini2016brain}. We next discuss those identified brain regions roughly following this trajectory.

\underline{\textit{The hippocampus and entorhinal cortex.}}
The hippocampus is a major component of the human brain located in the medial temporal lobe, and is functionally involved in response inhibition, episodic memory, and spatial cognition. Hippocampal atrophy is the best established and validated biomarker across the entire disease spectrum \citep{jack2011steps}. The entorhinal cortex also locates in the medial temporal lobe. It connects the neocortex and the hippocampus that receives information from the neocortex and projects to the hippocampus through the perforant pathway \citep{insausti1995human}. It has been consistently reported that, compared to the healthy controls, entorhinal atrophy was observed in the MCI patients, and more severe atrophy in the AD patients \citep{pini2016brain}. The hippocampus and entorhinal cortex, as well as the anatomically related parahippocampal and perirhinal cortices, are parts of the medial temporal lobe memory system. Impairments of this system are responsible for the deficit in episodic memory, and are early hallmark of AD \citep{nadel2011update}.

\underline{\textit{The lateral temporal, parietal, and frontal cortex.}}
The gray matter loss in the lateral temporal cortex, dorsal parietal, parietal angular and frontal cortex occurs during the progression from incipient to mild AD. During this period, cognitive deficits have been observed in both memory and non-memory domains, including language, visuo-spatial and executive function \citep{frisoni2009vivo}. Moreover, a higher amount of tau deposition has been observed in the middle temporal cortex, fusiform gyrus, and entorhinal cortex \citep{schultz2018widespread}.
The fusiform gyrus is critical in facial recognition. Alterations of gene expression specific to the fusiform gyrus were discovered in AD patients \citep{ma2020fusiform}.
The left middle temporal gyrus is related to the recognition of known faces and accessing word meaning while reading \citep{acheson2013stimulating}.
The precuneus, a hub of the default mode network, has been found to be related to episodic memories \citep{sadigh2014different}.
Atrophy in the entorhinal cortex, fusiform, middle temporal gyrus, precuneus, and precentral has been noted in AD \citep{parker2018cortical}.
The association between atrophy in the insular cortex and memory deficits in AD has been reported too \citep{lin2017insula}.

\underline{\textit{The lateral ventricles.}}
The ventricles are one of the interests in brain atrophy research as the volumetric measurement is robust to automatic segmentation due to the sharp contrast between the CSF in the ventricles and surrounding tissue in T1-weighted images. Thus, as a complement metric of hemispheric atrophy rates, enlargement in the lateral ventricles is an important marker of AD progression \citep{kruthika2019multistage}.

\underline{\textit{The cerebellum.}}
The cerebellum is involved in cognition and emotion and communicates with cerebral cortices in a topographically organized manner. Based on existing evidence of cerebellar modulation of cognition and emotion, it was hypothesized that there exists cerebellar contribution to the cognitive and neuropsychiatric deficits in AD. However, more research is required to validate the hypothesis and to understand cerebrocerebellar interactions in AD pathology~\citep{jacobs2018cerebellum}.

\section{Simulation Study}
\label{sec:sim}

We complement our data analysis with some additional simulation studies to further examine the empirical performance of the proposed method. 

We generate $\bX_{i}\in\mathbb{R}^{r}$ ($i=1,\dots,n$) from a multivariate normal distribution with mean zero and a covariance matrix whose eigenvalues exponentially decay. After applying PCA, we obtain $\tilde{\bX}\in\mathbb{R}^{n\times q}$, where $q$ is chosen such that the top $q$ PCs account for over $80\%$ of total data variation. We then generate $\bM$ and $\bY$ following model \eqref{eq:model} given $\tilde{\bX}$. We set $5\%$ of the path effects to be nonzero. We consider two sets of data dimension, $r=100, p=100$, and $r=350, p=150$, the latter of which has a similar data dimension as in the ADNI dataset. We also consider three sample sizes, $n=100, 500, 1000$. 

Table~\ref{table:sim_est} presents the estimated total indirect effects and the indirect effects of the top six PCs, and Table~\ref{table:sim_sens} presents the estimated number of PCs and the sensitivity and specificity of the identified nonzero path effects. Among all cases, the estimated number of PCs is 6, which agrees with the truth. From the tables, we observe that the proposed method  achieves a competitive performance, and the performance improves, with a lower estimation error and a higher selection accuracy, as the sample size increases.

\begin{table}
  \caption{\label{table:sim_est}The estimation bias and mean squared error (MSE) of estimating the total indirect effect and indirect effect of top PCs.}
  \begin{center}
    \begin{tabular}{c c l r r r c r r c r r c}
      \hline
      & & & & \multicolumn{2}{c}{$n=100$} && \multicolumn{2}{c}{$n=500$} && \multicolumn{2}{c}{$n=1000$} \\
      \cline{5-6}\cline{8-9}\cline{11-12}
      \multirow{-2}{*}{$r$} & \multirow{-2}{*}{$p$} & & \multicolumn{1}{c}{\multirow{-2}{*}{Truth}} & \multicolumn{1}{c}{Bias} & \multicolumn{1}{c}{MSE} && \multicolumn{1}{c}{Bias} & \multicolumn{1}{c}{MSE} && \multicolumn{1}{c}{Bias} & \multicolumn{1}{c}{MSE} \\
      \hline
      & & Total & -20 & 9.030 & 128.080 && -1.827 & 16.593 && -0.921 & 13.508 \\
      & & PC1 & -12 & 7.172 & 61.226 && 0.033 & 1.843 && 0.035 & 2.883 \\
      & & PC2 & 0 & -0.883 & 15.244 && -1.410 & 9.968 && -0.856 & 4.364 \\
      & & PC3 & -8 & 3.205 & 20.100 && -0.168 & 2.531 && -0.032 & 2.248 \\
      & & PC4 & 0 & -0.477 & 11.943 && -0.272 & 4.242 && -0.067 & 1.963 \\
      & & PC5 & 0 & -0.049 & 5.168 && -0.078 & 1.328 && 0.002 & 0.611 \\
      \multirow{-7}{*}{$100$} & \multirow{-7}{*}{$100$} & PC6 & 0 & 0.085 & 2.809 && 0.054 & 0.794 && 0.019 & 0.370 \\
      \hline
      & & Total & 8 & -6.317 & 80.520 && -1.164 & 37.593 && -1.284 & 19.271 \\
      & & PC1 & -8 & 6.461 & 54.644 && 0.723 & 13.823 && -0.511 & 3.616 \\
      & & PC2 & 12 & -9.528 & 102.378 && -0.933 & 20.122 && 0.816 & 3.045 \\
      & & PC3 & 4 & -3.446 & 27.204 && -1.398 & 10.500 && -1.740 & 7.489 \\
      & & PC4 & 0 & -0.093 & 8.015 && 0.147 & 2.595 && -0.033 & 1.403 \\
      & & PC5 & 0 & 0.153 & 5.445 && 0.060 & 1.615 && 0.066 & 0.781 \\
      \multirow{-7}{*}{$350$} & \multirow{-7}{*}{$150$} & PC6 & 0 & 0.137 & 6.489 && 0.238 & 1.054 && 0.117 & 0.750 \\
      \hline
    \end{tabular}
  \end{center}
\end{table}
\begin{table}
  \caption{\label{table:sim_sens}The estimated number of PC (and the standard error, SE) in the PCA step and sensitivity and specificity of identifying paths with a nonzero path effect.}
  \begin{center}
    \begin{tabular}{c c c c c c}
      \hline
      $r$ & $p$ & $n$ & \# PC (SE) & Sensitivity & Specificity \\
      \hline
      && $100$ & 6.03 (0.17) & 0.84 & 0.53 \\
      && $500$ & 5.21 (0.41) & 1.00 & 0.89 \\
      \multirow{-3}{*}{$100$} & \multirow{-3}{*}{$100$} & $1000$ & 6.26 (0.44) & 1.00 & 0.91 \\
      \hline
      && $100$ & 5.99 (0.10) & 0.80 & 0.57 \\
      && $500$ & 6.00 (0.00) & 1.00 & 0.89 \\
      \multirow{-3}{*}{$350$} & \multirow{-3}{*}{$150$} & $1000$ & 6.00 (0.00) & 1.00 & 0.91 \\
      \hline
    \end{tabular}
  \end{center}
\end{table}

\section*{Acknowledgments}


Data collection and sharing for this project was funded by the Alzheimer's Disease Neuroimaging Initiative (ADNI) (National Institutes of Health Grant U01 AG024904) and DOD ADNI (Department of Defense award number W81XWH-12-2-0012). ADNI is funded by the National Institute on Aging, the National Institute of Biomedical Imaging and Bioengineering, and through generous contributions from the following: AbbVie, Alzheimer's Association; Alzheimer's Drug Discovery Foundation; Araclon Biotech; BioClinica, Inc.; Biogen; Bristol-Myers Squibb Company; CereSpir, Inc.; Cogstate; Eisai Inc.; Elan Pharmaceuticals, Inc.; Eli Lilly and Company; EuroImmun; F. Hoffmann-La Roche Ltd and its affiliated company Genentech, Inc.; Fujirebio; GE Healthcare; IXICO Ltd.; Janssen Alzheimer Immunotherapy Research \& Development, LLC.; Johnson \& Johnson Pharmaceutical Research \& Development LLC.; Lumosity; Lundbeck; Merck \& Co., Inc.; Meso Scale Diagnostics, LLC.; NeuroRx Research; Neurotrack Technologies; Novartis Pharmaceuticals Corporation; Pfizer Inc.; Piramal Imaging; Servier; Takeda Pharmaceutical Company; and Transition Therapeutics. The Canadian Institutes of Health Research is providing funds to support ADNI clinical sites in Canada. Private sector contributions are facilitated by the Foundation for the National Institutes of Health (\url{www.fnih.org}). The grantee organization is the Northern California Institute for Research and Education, and the study is coordinated by the Alzheimer’s Therapeutic Research Institute at the University of Southern California. ADNI data are disseminated by the Laboratory for Neuro Imaging at the University of Southern California.


\bibliographystyle{apalike}
\bibliography{ref-mexp}

\end{document}